\documentclass[12pt]{article}
\usepackage{epsfig}

        \newdimen\eqskip
        \newdimen\txtskip
        \baselineskip=\txtskip
        \newdimen\mysep                
        \newdimen\hmysep
\def\d{\mathrm{\;d}}
\def\Mvariable#1{#1}
\def\MathBegin#1#2{}
\def\MathEnd#1{}
\def\noalign#1{}

\let\ShortRightArrow\rightarrow
\def\inlinemio#1{$ #1$}
\def\inlinebig#1{\begin{equation} #1 \end{equation}}
\def\nn{\nonumber}

\def\be{\begin{equation}}
\def\ee{\end{equation}}
\def\ba{\begin{eqnarray}}
\def\ea{\end{eqnarray}}

\def\beq{\begin{equation}} \def\eeq{\end{equation}} \def\bea{\begin{eqnarray}}
\def\eea{\end{eqnarray}}
\def\bq{\begin{quote}} \def\eq{\end{quote}}

\def\gappeq{\mathrel{\rlap {\raise.5ex\hbox{$>$}} {\lower.5ex\hbox{$\sim$}}}}

\def\lappeq{\mathrel{\rlap{\raise.5ex\hbox{$<$}} {\lower.5ex\hbox{$\sim$}}}}

\def\CellGroup{\bgroup}
\def\endCellGroup{\egroup}
\begin{document}
\begin{titlepage}
\nopagebreak       {\flushright{
        \begin{minipage}{5cm}
        LPT-ORSAY 00 36\\
         March 00\\
        \end{minipage}        }

}
\vfill 
\begin{center} 
{\LARGE  { \bf \sc  How to  Integrate Divergent
Integrals:  \\[0.5cm]   a  pure  numerical  approach   to  complex  loop
calculations}} 
\vfill 
{\bf F. CARAVAGLIOS }\\[1cm]
{\it  LPT,   B\^at.  210,  Univ.  Paris-Sud,   F-91405,  Orsay,  France}
\end{center}  
\nopagebreak
\vfill 
\begin{abstract}
  Loop
calculations involve the evaluation of divergent integrals.
  Usually\cite{'tHooft:1972fi} one
computes them  in a number of  dimensions different than 
 four  where the
integral is convergent and then one performs the analytical continuation
and considers the Laurent expansion in powers of $\varepsilon =n-4$.  In
this paper we  discuss a method to extract  directly all coefficients of
this expansion by means of concrete and well defined integrals in a five
dimensional  space. We  by-pass  the formal  and  symbolic procedure  of
analytic continuation;  instead we can  numerically compute
the   integrals  to   extract  directly both   the  coefficient   of   
the  pole $1/\varepsilon$  and   the  finite  part.  
 \end{abstract} 
 \vskip  1cm
LPT-ORSAY 00-36\hfill \\
 March 00 \\
\vfill 
\end{titlepage}
 \section*{Introduction} 
   Feynman    diagrams
computations  are often  a lengthy  and  hard task.   The complexity  of
typical computations grows as a factorial  with the number of external legs
and/or loops involved in the  processes.  In the Standard Model the main
difficulties arise  from QCD corrections  and the production of  jets in
the final states;  but it is not difficult to  imagine extensions of the
Standard  Model,  in  which  this  problem  could  seriously  limit  our
capabilities of  studying and  understanding new physical  phenomena; in
particular  if they  are  involved (in  addition to  \inlinemio{{{\alpha
}_s}}  )   by  new   rather  strong  couplings,   which  give   rise  to
multiparticle/multiloop  amplitudes.  This  issue motivates  a  rich and
increasing  activity. Considerable  progresses have  been achieved 
 in the  past \cite{Mangano:1991by}:
  new methods
inspired by  string theory [3-5],
  helicity 
amplitudes [6-9],
 different recursive 
algorithms to calculate QCD dual amplitudes\cite{Berends:1988me},
 or the scattering
matrix elements of generic  processes with arbitrary initial/final 
states\cite{Caravaglios:1995cd}.
  Very useful approximations have been
 proposed [12-16],
 when the exact matrix 
elements are unknown. 
In loop computations interesting 
simplifications [17-26]
 can be used.
 
While analytical methods can provide  us in some specific processes with
very simple results, a general approach for generic lagrangian and final
states cannot avoid the use  of the computer power.  A possible strategy
is to implement the Feynman rules into some automatic code with the help
of  some  familiar  packages  for  symbolic  manipulation  of  algebraic
formulae.   However  this  way  to  proceed  becomes  ineffective,  when
the computer has to manage very complex formulae (as often it is the case).

 This  suggests a second  strategy which  tries exploiting  the computer
 power with pure numerical algorithms and thus avoiding lengthy symbolic
 manipulation.   In   the  recent   past  these  algorithms   have  been
 successfully used for the  computation of several tree level amplitudes
 both in QCD\cite{Draggiotis:1998gr,caravaglios:1999} and in electroweak
 processes [29-31].
 In this paper we discuss the
 possibility to apply pure numerical techniques\cite{Soper:1999xk} 
in loop calculations.

 The  problem of  performing loop  computations  can be  divided in  two
 parts.  The  first part is similar  to tree level 
 computations\cite{Caravaglios:1995cd}: from a
 given lagrangian one has to produce an explicit function of virtual and
 real momenta which is equivalent to the sum of all the Feynman diagrams
 contributing  to the process.   Once this  task is  accomplished, loops
 calculations  have an  additional problem  with respect  to  tree level
 ones: integrals  over virtual loop momenta are  affected by ultraviolet
 and/or   infrared   divergences.    These   divergences   require   the
 introduction of a regularization  procedure which usually obliges us to
 perform all virtual loop  integrations in an analytical and essentially
 symbolic fashion (while in tree level calculations one usually performs
 the integration over the real momenta by numerical montecarlo methods).
 This  is a  serious  difficulty  in any  numerical  approach.  Here  we
 address this last  issue. We will define a  {\it numerical\/} procedure
 to  perform   virtual  loop   integrations  even  when   these  include
 divergences.

 Let us consider a function \inlinemio{f[{l_{\mu }},{p_{\mu }} ]} of the
 external  momenta \inlinemio{{p_{\mu }}  } and  of the  virtual momenta
 \inlinemio{{l_{\mu }}}.   The $n$ dimensional  
integration\cite{'tHooft:1972fi} is performed
 in a region of complex values  of $n$ where the integral is convergent;
 then one considers  the analytical continuation to values  of $n$ where
 the integral  is not  well defined.   In general for  $n \simeq  4$ the
 analytical    continuation   can    have    a   pole    $1/\varepsilon$
 ($\varepsilon=n-4$).    We  can  use   the  Laurent   expansion  around
 $\varepsilon \simeq  0$ to define  some functionals $I_k[f]$  such 
that\footnote{At one loop.}
 \inlinebig{\label{eq1}\int               {\d^n}l              f[{l_{\mu
 }}]={I_{-1}}[f]{{\varepsilon   }^{-1}}  +{I_0}[f]+{I_1}[f]{{\varepsilon
 }^1}+{I_2}[f]{{\varepsilon  }^2}+\ldots  }  Since the  $n$  dimensional
 integration is well defined for  any function $f$, also the functionals
 $I_k[f]$  are  unambiguously defined  from  equation (\ref{eq1}).   For
 instance,  $I_0[f]$  coincides   with  the  ordinary  four  dimensional
 integral \inlinebig{\label{eq2}\int {\d^4} \bar l f[{l_{\mu }}]} if $f$
 is a convergent function.  While  the left-hand side of the (\ref{eq1})
 is essentially a formal definition which does not lead us to an obvious
 numerical integration procedure, each  $I_k[f]$ can be written in terms
 of well defined and convergent integrals.  In the following we show how
 to  build   explicit  integral  representations   for  the  functionals
 $I_k[f]$.  In  the next sections, we explain  two different definitions
 (even  if equivalent)  of the  $I_k[f]$; in  the final  part  some very
 simple  numerical examples  will be  given  in order  to emphasize  the
 practical purposes  of the method.  \section*{Method  of Integration A}
 Let us assume that we have  to perform an integral in $n>4$ dimensions.
 The virtual momentum $l_\mu$ has  $n$ components.  We call $\bar l_\mu$
 the components  with $\mu\leq4$ and $\tilde l_\mu$  the components with
 $\mu>4$. Then  we can rewrite the  integral \inlinebig{\label{eq3} \int
 {\d^n}l  f[{l_{\mu  }},{p_{\mu }}  ]=  \int  {\\d^4}  \bar l  f[{l_{\mu
 }},{p_{\mu  }} ]{\d^{\varepsilon  }} \tilde  l  = \int  {\\d^4} \bar  l
 f[{l_{\mu  }},{p_{\mu }}  ]{\tilde  l^{\varepsilon-1 }}\d  \tilde l  \d
 \Omega_\varepsilon} $ \d \Omega_\varepsilon$  is the solid angle in the
 subspace  ($\mu>4$)   of  dimension  $\varepsilon$   orthogonal  to  the
 four-vector $\bar  l_\mu$.  Note that  $ f[{l_{\mu }},{p_{\mu }}  ]$ is
 invariant under rotations in  this subspace, since the external momenta
 do  not have  components $\mu>4$.  We  can exploit  this invariance  as
 follows.  First we  rotate the vector $\tilde l_\mu$  such that $\tilde
 l_5 \ne  0$ and  $\tilde l_n  = 0$ for  any $n>5$.   Thus we  can write
 $l_\mu=\big({\bar     l_1},{\bar     l_2},\bar    {l_3},\bar     {l_4},
 {\sqrt{t}}\big)$                                                    with
 ${t={{({l_5})}^2}+{{({l_6})}^2}+...+{{({l_n})}^2}}$  and  omit all  the
 components  with $\mu>5$.  We  have reduced  our $n$  dimensional space
 into a five dimensional space.   Second, we can perform the integral in
 $\d   \Omega_\varepsilon$    and   we   obtain    from   eq.(\ref{eq3})
 \inlinebig{\label{eq4}    \frac{{{\pi    }^{\varepsilon    /2}}}{\Gamma
 \big[\frac{\varepsilon   }{2}\big]}    \int   _{0}^{\infty   }\Big(\int
 {\d^4}\bar  l  f[{l_{\mu }},{p_{\mu  }}  ]\Big) {t^{\varepsilon  /2-1}}
 \Mvariable{\d t }.}  Now  the integral in \inlinemio{\Mvariable{\d t }}
 can   be    integrated   by   parts   \inlinebig{\label{eq5}\frac{{{\pi
 }^{\varepsilon  /2}}}{\Gamma   \big[\frac{  \varepsilon  }{2}\big]}\int
 _{0}^{\infty  }   \frac{4  {t^{1+\frac{\varepsilon  }{2}}}}{\varepsilon
 (2+\varepsilon      )}\bigg(     \int      {\d^4}      \bar     l\frac{
 {\d^2}}{{{\Mvariable{\d  t  }}^2}}   f[{l_{\mu  }},{p_{\mu  }}  ]\bigg)
 \Mvariable{\d t }.}  The function  $f$ depends on $t$ through the fifth
 component of  $l_\mu$.  The function \inlinemio{{t^{1+\frac{\varepsilon
 }{2}}}}  has  a  cut in  the  positive  real  axis of  \inlinemio{t}  (
 \inlinemio{\Mvariable{Re}(t)>0} and \inlinemio{\Mvariable{Im}(t)=0} ) ,
 in     fact     the     rotation     \inlinemio{t\ShortRightArrow     t
 {e^{\Mvariable{i2\pi  }}}  }  gives  \inlinemio{{t^{1+\frac{\varepsilon
 }{2}}}\ShortRightArrow  {t^{1+\frac{\varepsilon  }{2}+\Mvariable{i  \pi
 \varepsilon  }}}}.   Therefore  we  can apply  the  following  identity
 \inlinebig{\label{eq6}  \int   _{0}^{\infty  }  {t^{1+\frac{\varepsilon
 }{2}}}f[t]   \Mvariable{\d  t   }=  \frac{1}{\big(1-{e^{\Mvariable{i\pi
 \varepsilon   }}}\big)}    \oint   {t^{1+\frac{\varepsilon   }{2}}}f[t]
 \Mvariable{\d  t  }} where  the  contour  of  the last  integration  in
 \inlinemio{t} must contain all the poles and singularities lying in the
 complex \inlinemio{t} plane and must avoid the cut on the positive real
 axis  (see figure  1). Then the  integral  (\ref{eq5}) above  becomes
 \inlinebig{\label{eq7}\frac{1    }{(1-{e^{\Mvariable{i\pi   \varepsilon
 }}})}\frac{{{\pi   }^{\varepsilon  /2}}}{\Gamma  \big[\frac{\varepsilon
 }{2}\big]}\oint   \frac{4  {t^{1+\frac{\varepsilon  }{2}}}}{\varepsilon
 (2+\varepsilon    )}   \bigg(    \int   {\d^4}\bar    l\frac{   {\d^2}}
 {{{\Mvariable{\d t }}^2}} f[{l_{\mu }},{p_{\mu }} ]\bigg) \Mvariable{\d
 t.   }   }   Now   we   can   make   the   expansion   in   powers   of
 \inlinemio{\varepsilon } (the Euler  constant and $\log(\pi)$ have been
 re-absorbed  into  a  redefinition  of   the  scale  $\mu$  as  in  the
 $\overline{MS}$) and we  obtain the final result \inlinebig{\label{eq8}
 \oint \frac{i t}{2 \pi }\Big(\frac{2}{\varepsilon }+(-1+\log [-t])\Big)
 \; \bigg(  \int {\d^4} \bar  l \frac{ {\d^2}}{{{\Mvariable{\d  t }}^2}}
 f[{l_{\mu  }},{p_{\mu }}  ]\bigg) \Mvariable{\d  t.  } }  In the  above
 expression each  integral is now convergent,  due to the  action of the
 derivative $ \d^2/ \d t^2$ appearing inside the integral in $ \d^4 \bar
 l$.   Thus  one can  separately  calculate  the  singular and  the  non
 singular part, simply evaluating well defined and convergent integrals.
 We can  check the result  above in a  specific example and see  how the
 above formula can be used for practical calculations.

Suppose that  we have to  evaluate the following divergent  integral \be
\label{eq9} \int {\d^n}l\frac{{l^2}}{{l^2}+m^{2}}.   \ee In our approach
$l_\mu$ is not  a $n$ dimensional object but it is  a more concrete five
dimensional  vector  with  components  \inlinemio{\big({\bar  l_1},{\bar
l_2},\bar   {l_3},{\bar   l_4},   {\sqrt{t}}\big)};   thus   $l^2=l_\mu
l^\mu=\bar  l^2 +t$  with \inlinemio{{\bar  l^2}} equal  to  the squared
length   of  the   real  four-vector   \inlinemio{\big({\bar  l_1},{\bar
l_2},{\bar l_3},\bar {l_4}\big)} and  \inlinemio{t} a complex number. If
we are interested in the finite part, we can apply the non singular term
in  equation  (\ref{eq8})  and  we  have  \inlinebig{\label{eq10}  \oint
\frac{i t}{2  \pi }(-1+\log [-t])\bigg(\frac{  {\d^2}}{{{\Mvariable{\d t
}}^2}}\frac{{\bar   l^2}+t}{{\bar  l^2}+m^{2}+t}   \bigg)  \hspace{2.em}
{\d^4}   \bar   l  \Mvariable{\d   t.   }\\   }   After  the   derivative
\inlinemio{\frac{  {\d^2}}{{{\Mvariable{\d  t  }}^2}}} the  integral  in
$\d^4 \bar l$ is convergent\footnote{We  consider only value of $t$ with
non  zero  imaginary  part.  See  the  discussion at  the  end  of  this
section.},  and this  yields  \be \label{eq11}  \oint  \frac{i t}{2  \pi
}(-1+\log   [-t])\bigg(   -\frac{{{\pi   }^2}   m^{2}}{t+m^{2}}   \bigg)
\Mvariable{\d t } \ee The integral over the contour in the complex plane
is   equal    to   the   residue    at   the   pole   $t    \sim   -m^2$
\inlinebig{\label{eq12}-{{\pi  }^2}  \big(-1+\log  [m^{2}]\big)  m^{4}.}
Our final  result is correct.  The  advantage of the  procedure above is
clear: eqs.(\ref{eq10}-\ref{eq11}) are well  defined and concrete
 expressions, any
step can  be done  in a pure  numerical way.  Note that the  integral in
$\d^4 \bar  l$ must  be done  after the derivative  $\d^2 /\d  t^2$, and
before the integral  in $\d t$; otherwise the integral  in $\d^4 \bar l$
would be  non convergent.  In practice,  if the above  procedure is done
numerically, one should replace the integral in $\d t$ with a sum over a
finite set of points $t_i$; for  the convergence of all integrals, it is
enough to choose  the $t_i$ in this set with a  non zero imaginary part.
\section*{Method  of  Integration  B}   The  formula  (\ref{eq8})  is  a
transparent and compact expression, which makes manifest some properties
of the  dimensional regularization;  for instance, the  invariance under
the translations $\bar l_\mu \rightarrow \bar l_\mu + p_\mu$ is obvious,
since  it  comes  directly  from  the translational  invariance  of  the
integral in $\d^4 \bar l$,  which is convergent (after the derivative in
$\d^2  /  \d  t^2$).   However  in certain  numerical  calculations  the
(\ref{eq8}) could be not efficient.  We discuss a different method which
is  more powerful  in  practical  calculations, even  if  it requires  a
slightly more involved formula.

There are several  different ways of rewriting the  (\ref{eq8}); each of
them depending on the way we choose to parameterize the virtual momentum
$l_\mu$.  Here  we do not  intend to make  an exhaustive study,  we will
simply  describe  a  quite  general  procedure to  obtain  well  defined
integral representations for the $I_k$ in (\ref{eq1}).

First  we observe that  each \inlinemio{{I_k}[f]}  is a  linear operator
  \inlinebig{\label{eq13}{I_k}[{f_1}+{f_2}]={I_k}[{f_1}]+{I_k}[{f_2}]}
  \inlinebig{\label{eq14}  {I_k}[\lambda  f]=\lambda  {I_k}[f].}  It  is
  quite natural  to think  that the linearity  (\ref{eq13}-\ref{eq14}) implies
  that the $I_k$  are authentic integrals or sum  of integrals.  In fact
  we  can   introduce  a  simple   trick  to  build   concrete  integral
  representations for generic linear functionals.

Consider the space of functions  which admits a Laurent expansion around
a   point   \inlinemio{{t_0}}.   They   are   defined   by   a  set   of
\inlinemio{{a_n}} through the expansion\footnote{We also assume that the
series  converges   strongly  enough   to  justify  the   steps  below.}
\inlinebig{\label{eq15}f[t]=\sum          _{n=-\infty          }^{\infty
}{a_n}{{(t-{t_0})}^n}.}  A linear  functional \inlinemio{I[f]} acting on
this       set       of       functions       can       be       written
\inlinebig{\label{eq16}I\big[f\big]=I\bigg[  \sum  _{n=-\infty }^{\infty
}{a_n}       {{(t-{t_0})}^n}\bigg]=\sum       _{n=-\infty      }^{\infty
}{a_n}I[{{(t-{t_0})}^n}]= \sum _{n=-\infty }^{\infty } {a_n}{b_n}} where
\inlinemio{{b_n}=} \inlinemio{I[{{(t-{t_0})}^n}]} is a real (or complex)
number.    Then  we   can  use   the   identity  \inlinebig{\label{eq17}
{{\oint}_C}\,\frac{1}{2                                               \pi
i}\frac{{{(t-{t_0})}^n}}{{{(t-{t_0})}^{k+1}}}\Mvariable{\d  t }={{\delta
}_{\Mvariable{nk}}}}  for any \inlinemio{n,k=0,\pm  1,\pm 2,\ldots  } if
the  integral contour  C is  a  small circle,  in the  complex plane  of
\inlinemio{t},   around   the   point  \inlinemio{{t_0}}.    Using   the
(\ref{eq17}),  the  (\ref{eq16}) can  be  written  \ba \label{eq18}  \nn
I[f]&=&\frac{1}{2   \pi    i}\oint   \big(\sum   _{n=-\infty   }^{\infty
}{a_n}{{(t-{t_0})   }^n}\big)    \Big(\sum   _{k=-\infty   }^{\infty   }
\frac{{b_k}}{{{(t-{t_0})}^{k+1}}}\Big)\Mvariable{\d   t    }   =\\   \nn
&=&\frac{1}{2  \pi  i}\oint  f[t]  w[t]\Mvariable{\d  t }  \\  \ea  with
\inlinebig{\label{eq19}w[t]=\sum     _{k=-\infty     }^{\infty    }\Big(
\frac{{b_k}}{{{(t-{t_0})}^{k+1}}}\Big).}     This    is   an    integral
representation of the  linear functional $I$.  This simple  trick can be
used  to build  explicit and  compact integral  representations  for any
linear functional, whose  action on each term of  a Laurent expansion is
known.
 
Let  us  apply  it  to  our  problem.  We  start  defining  the  angular
integration in $n=4+\varepsilon$ dimensions.  The analytical integration
of  any tensor  with $k$  (even) indices  and constructed  with  the $n$
dimensional  vector  $l_\mu$  yields\footnote{Here  we assume  that  the
overall factor  $ {\pi^{\frac{\varepsilon}{2}}/{ \Gamma[ \frac{n}{2}]}}$
has  been  re-absorbed  into  a  redefinition of  the  scale  $\mu$,  as
prescribed by the $\overline{MS}$ scheme.}  \inlinebig{\label{eq20} \int
{l_{\bar \mu  }}{l_{\bar \nu }}{l_{\bar \alpha }}  \ldots {l_{\bar \beta
}}   {l_{\bar   \rho    }}{l_{\bar   \sigma   }}{{\Mvariable{\d   \Omega
}}_{4+\varepsilon    }}=2    {\pi^2}    \frac{   (2+\varepsilon    )!!}{
(2+k+\varepsilon  )!!}\big(  {g_{\Mvariable{   \bar  \mu  \bar  \nu  }}}
{g_{\Mvariable{  \bar \rho  \bar \sigma  }}}\ldots  {g_{\Mvariable{ \bar
\alpha  \bar  \beta }}}+\ldots  +{g_{\Mvariable{  \bar  \mu \bar  \alpha
}}}{g_{\Mvariable{  \bar  \rho  \bar \nu  }}}\ldots{g_{\Mvariable{  \bar
\beta \bar \sigma }}}\big)\ {{l^{2(k/2)}}} } \inlinemio{ {{\Mvariable{\d
\Omega  }}_{4+\varepsilon }}={{\sin  }^{2+\varepsilon  }}[\gamma ]\ldots
{{\sin }^2}[\theta ]{{\sin}}[\phi]\Mvariable{\d \gamma }\, \Mvariable{\d
\theta }\, \Mvariable{\d \phi }\, \Mvariable{\d \delta }\, ~~~ (0<\gamma
,\theta,\dots ,\phi <\pi ;0<\delta <2\pi )  } is the solid angle.  It is
understood that the  indices ($\bar \mu $, $\bar  \nu \ldots$) above are
contracted  with some  external momenta  and/or other  tensorial indices
(like the spin  of the external particles), since  the scattering matrix
is  lorentz invariant.  Thus\footnote{We  remind that  only $l_\mu$  has
components $ >4$, and factors  $l_\mu l^\mu=l^2$ are not relevant for the
angular integration.}  only indices with  $\bar \mu \leq 4$ are relevant
in our problem.  This has been  emphasized by the small bar on the Greek
indices.

We look  for an integral  definition which reproduces  the (\ref{eq20}).
To begin  the procedure we parameterize  $l_\mu$ with an  array of five
components\footnote{In  practice this means  that any  vector (including
external momenta etc.), must be  written as an array of five components,
but we  keep in  mind that  only \inlinemio{{l_{\mu }}}  has a  non zero
fifth  component.}  \inlinebig{\label{eq21}  {l_{\mu  }}\rightarrow t  \
\Big(x \cos  [\theta ],x \sin  [\theta ] \cos  [\phi ],x \sin  [\theta ]
\sin [\phi ] \cos [\delta ], x  \sin [\theta ] \sin [\phi ] \sin [\delta
],{\sqrt{1-{x^2}}} \Big) } $\theta $ $\phi $ and $\delta $ are the three
angles in the space of four dimensions.  \inlinemio{{t}} and \inlinemio{
x} are  two complex  variables which will  be integrated over  a complex
contour  as  explained  below.   The  need  of  the  auxiliary  variable
\inlinemio{ x}  will be  clear in a  moment.  Suppose that  we integrate
over  the  four  dimensional  solid angle  \inlinemio{{  {{\Mvariable{\d
\Omega  }}_4}}={{\sin  }^2}[\theta  ]{{\sin }^{  }}[\phi  ]\Mvariable{\d
\theta }  \Mvariable{\d \phi }  \Mvariable{\d \delta }} all  the tensors
built with the array (\ref{eq21}) in place  of an $n$ dimensional $l_\mu$.
We  get \be  \label{eq22} \int  {l_{\bar \mu  }}{l_{\bar  \nu }}{l_{\bar
\alpha  }} \ldots  {l_{\bar \beta  }} {l_{\bar  \rho }}  {l_{\bar \sigma
}}{{\Mvariable{\d \Omega  }}_4}= 2{{\pi }^2} \frac{ (2)!!}{  ( 2+k )!!}(
{g_{\Mvariable{  \bar \mu  \bar  \nu }}}{g_{\Mvariable{  \bar \rho  \bar
\sigma  }}}\ldots{g_{\Mvariable{  \bar   \alpha  \bar  \beta  }}}+\ldots
+{g_{\Mvariable{ \bar \mu \bar  \alpha }}}{g_{\Mvariable{ \bar \rho \bar
\nu  }}}\ldots{g_{\Mvariable{ \bar  \beta \bar  \sigma  }}}){t^k} {x^k.}
\ee Comparing  this result with  the (\ref{eq20}) we see  that tensorial
structure is identical,  but the overall factor is not  correct. The term of
order $\varepsilon^0$  can be  obtained  simply setting $x=1$  and $t^2=l^2$.
Instead for  the term  of order  $\varepsilon $, it  would be  enough to
replace $x^k$ with  a suitable factor\footnote{ The function  $ \psi$ is
the well known derivative $\d/\d x\; \log[\Gamma[x]]$; it comes out when
we  expand the factorial  in the  (\ref{eq20}) and  take only  the first
order  in  $\varepsilon$.   } 
 \inlinebig{\label{eq23}  {x^k}\rightarrow
b_k=\frac{\varepsilon}{2}  (1-\gamma_E -\psi  [2+k/2])}  for any  even
\inlinemio{k}.

This  can be  achieved by  means of  a linear  functional $I$  such that
 $I[x^k]=b_k$.     In    fact,     following     the    recipe     above
 (eqs.(\ref{eq15})-(\ref{eq19}))  we   can  build  the   integral  below
 \inlinebig{\label{eq24}  I[{f}]=\frac{\varepsilon}{2\pi  i}\oint {{\sum
 }_{k=\Mvariable{even}}}       \frac{b_k}{x^{k+1}}       {f[x]}\d      x
 =\frac{\varepsilon}{2\pi        i}\oint        \frac{x+{x^3}       \log
 \big[1-\frac{1}{{x^2}}\big]}{2  (-1+{x^2})}  {f[x]}\d  x  }  where  the
 integral contour  is a circle around the  singularity \inlinemio{ x\sim
 0}.  The  integrand above  has a cut  in the  segment of the  real axis
 between -1  and 1.  It is  easy to see that  in the limit of  a path of
 integration very close  to this cut, the logarithm  can be approximated
 by its  imaginary part.  The  real part does  not contribute to  the full
 integral, and  the logarithm can be  replaced by $\pm i  \pi $.  Taking
 also into  account that $f[x]=f[-x]  $ we can rewrite  the (\ref{eq24})
 \inlinebig{\label{eq25}I[{f}         ]=         \varepsilon        \int
 _{0}^{1}{{\bigg(\frac{{x^3}}{   1-{x^2}}\bigg)}_+}  {f[x]}\Mvariable{\d
 x.}  }  Here  the  notation  \inlinemio{ {{(\ldots  )}_+}}  means  that
 \inlinebig{\label{eq26}\int    _{0}^{1}{{(h[x])}_+}   f[x]\Mvariable{\d
 x}=\int _{0}^{1}  h[x]\; (f[x]-f[1])\Mvariable{\d x}, }  and this makes
 the integral convergent near $x\sim 1$.

 Finally  we  are  able to  write  a  compact  formula for  the  angular
 integration in  $4+\varepsilon$ dimensions \inlinebig{\label{eq27} \int
 f[{l_{\mu }}]{{\Mvariable{\d\Omega  }}_{4+\varepsilon }}= \hspace{1.em}
 {{\Big(\int     f[{l_{\mu     }}]{{\Mvariable{\d\Omega     }}_{4     }}
 \Big)}_{x=1}}+\varepsilon  \bigg(\int  _{0}^{1}  \int  f[{l_{\mu  }}]{{
 \bigg(\frac{{x^3}}{      1-{x^2}}\bigg)}_+}\Mvariable{\d      x}     {{
 \Mvariable{\d\Omega }}_{4 }}\bigg)+ {{\varepsilon }^2}(\ldots ).  }

 The  last step  is to  define the  integration in  \inlinemio{\d l^2=\d
   t^2}.  Again  we follow the trick  (\ref{eq15})-(\ref{eq19}): we have
   to compute the $b_\sigma$ in  order to get the function $w[t^2]$.  By
   definition,  we  know  that   for  any  \inlinemio{\sigma  >2  }  \ba
   \label{eq28}b_\sigma&=&I\big[\frac{1}{(t^2+m^2)^\sigma     }\big]=\int
   {t^{2(1+\varepsilon               /2)}}               {{\Mvariable{\d
   t}}^2}\frac{1}{{{({t^2}+{m^2})}^{\sigma   }}}  =\nn  \\   &=&  \frac{
   {1}}{(-2+\sigma     )     (-1+\sigma    )}\frac{1}{{{({m^2})}^{\sigma
   -2}}}+\varepsilon   (\ldots   ),\\   \nn   \ea  for   $\sigma   $=1,2
   \ba\label{eq29}b_1&=&2    \frac{{{m}^2}}{\varepsilon   }+{m^2}   \log
   [{m^2}]+\varepsilon         (\ldots          )         \nn         \\
   \label{eq30}b_2&=&={-}\frac{2}{\varepsilon        }        -\big(\log
   [{m^2}]+1\big)+\varepsilon   (\ldots   )    \ea   and   for   $\sigma
   $=0,-1,-2,...    the  integral  is   zero  ($b_\sigma=0$).    If  the
   integration contour in the \inlinemio{{t^2}} complex plane is a small
   circle  around  the  pole  singularity  \inlinemio{{t^2}=-{m^2}}  and
   replacing  the   \inlinemio{{b_n}}  in  the   (\ref{eq19})  with  the
   \inlinemio{{b_{\sigma }}  } above  we get \be  \label{eq31} w[{t^2}]=
   \sum                _{\sigma=0}^{\infty               }{{b_{\sigma+1}
   \big({t^2}+{m^2}\big)}^{\sigma}}  =-{t^2} \bigg( \frac{2}{\varepsilon
   }+ \log  \big[-{t^2}\big]\bigg) \ee and  \inlinebig{\label{eq32} \int
   {t^{2(1+\varepsilon      /2)}}     {{\Mvariable{\d     t}}^2}f[{t^2}]
   =-\frac{2}{\varepsilon   }\Big(\frac{1}{2\pi   i}\oint  f[{t^2}]{t^2}
   {{\Mvariable{\d  t}}^2}\Big)-   \Big(\frac{1}{2\pi  i}\oint  f[{t^2}]
   {t^2}  \log \big[-{t^2}\big]{{\Mvariable{\d  t}}^2}\Big) +\varepsilon
   (\ldots ).  } The integral contour is closed, it must contain all the
   singularities    in    the   complex    plane    of   the    function
   \inlinemio{f[{t^2}]}, except  for the singularity in  $t^2=0$ and the
   cut for  real and positive  \inlinemio{{t^2}} in all  integrals where
   the logarithmic function appears (see for example figure 1).  We will
   comment on this contour of integration later on.

Combing the angular integration (\ref{eq27}) and the (\ref{eq32}) we get
 the full formula
\be   \label{eq33}    \int   f[{l_{\mu   }}]{\d^n}l=    \int   f[{l_{\mu
}}]{t^{2(1+\varepsilon      /2)}}\frac{{     \Mvariable{\d     t}}^2}{2}
{{\Mvariable{\d           \Omega           }}_{4+\varepsilon          }}
=\frac{I_{-1}}{\varepsilon}+I_{0}+I_{1} \varepsilon+\ldots \ee where \ba
\label{eq34bis}         I_{-1}\big[         f[{l_{\mu        }}]\big]&=&
{{\Big(\frac{1}{\Mvariable{   2  \pi   i   }}\oint  f[{l_{\mu   }}]{t^2}
{{\Mvariable{\d   t}}^2}{{\Mvariable{\d  \Omega   }}_4}  \Big)}_{x=1}}\\
\label{eq34}I_{0}\big[  f[{l_{\mu }}]\big]&=&{{\Big(\frac{1}{\Mvariable{
4 \pi  i }}\oint f[{l_{\mu  }}]{t^2}\log \big[-{t^2}\big]{{\Mvariable{\d
t}}^2}{{\Mvariable{\d \Omega  }}_4}\Big)}_{x=1}}+ \frac{1}{\Mvariable{ 2
\pi   i   }}\int   _{0}^{1}\oint  f[{l_{\mu   }}]{t^2}   {{\Mvariable{\d
t}}^2}{{\bigg(\frac{{x^3}}{    1-{x^2}}    \bigg)}_+}\Mvariable{\d    x}
{{\Mvariable{\d \Omega  }}_{4 }}. \nn\\ \ea Clearly  the integrals above
are well  defined and can  be evaluated numerically.\footnote{  Here all
integrals  are understood  to be  in Euclidean  space. In
some kinematical regions, the  integration (even if convergent) needs to
be   regularized   by  the   prescription   $m^2\rightarrow   m^2  +   i
\varepsilon$. Also for some infrared singularities the (\ref{eq34}) must
be rearranged differently or one  should use the method A. The practical
use of these methods in various realistic situations certainly demands a
much more  extensive discussion,  in this letter  we simply  present the
general  idea,  a  more  complete  and systematic  study  will  be  done
elsewhere  (see ref.\cite{caravaglios:2000}).   } 
For example  to
evaluate the first integral of $I_0$ in the expression (\ref{eq34}), the
contour of integration in $\d t^2$ can be chosen as in figure 1. Namely we
can divide the contour in three paths.   The path A is very close to the
cut of  the logarithm. This  function can be  replaced with $\pm  i \pi$
(the sign  depends if we are above  or below the cut).   This yields the
following integral  \be \label{eq35} P_1={{\Big(\frac{1}{\Mvariable{2 }}
\int_{\Lambda       _{\Mvariable{ir}}^{2}}^{\Lambda_{\Mvariable{uv}}^{2}}
f[{l_{\mu     }}]{t^2}{{\Mvariable{\d    t}}^2}{{\Mvariable{\d    \Omega
}}_4}\Big)}_{x=1}}. \ee
 \begin{figure}[h]  \center \vskip
0.6cm   \epsfig{figure=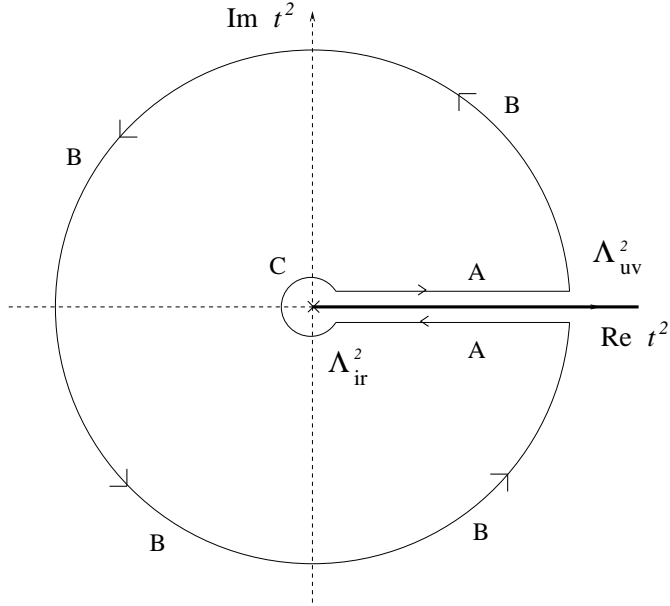,  height=   8cm}  \caption{Integration
contour in the $t^2$ complex plane.  The thick line in the positive real
axis represents the cut of the logarithmic function.  The path A is very
close  to this  cut  and the  logarithm  can be  replaced  there by  its
imaginary part. $\Lambda_{uv}^2$  and $\Lambda_{ir }^2$ are respectively
the  infrared and  ultraviolet  cut-off.}  
\end{figure}
  Note that this  is an ordinary  four dimensional
integral  with an  infrared and  an ultraviolet  cut-off: in  fact after
setting  $x=1$, the  array (\ref{eq21})  may  be regarded  as a  genuine
four-vector.  Then we have the paths  B and C, that are two circles with
radius   $\Lambda_{uv}^2$    and   $\Lambda_{ir}^2$   respectively   \be
\label{eq36} P_2={{\Big(\frac{1}{\Mvariable{ 4 \pi i }}\oint_B f[{l_{\mu
}}]{t^2}\log    \big[-{t^2}\big]{{\Mvariable{\d    t}}^2}{{\Mvariable{\d
\Omega }}_4}\Big)}_{x=1}}+ {{\Big(\frac{1}{\Mvariable{ 4 \pi i }}\oint_C
f[{l_{\mu          }}]{t^2}\log          \big[-{t^2}\big]{{\Mvariable{\d
t}}^2}{{\Mvariable{\d  \Omega }}_4}\Big)}_{x=1}}.   \ee  These integrals
can be  seen as  some counterterms which  cancel the  $\Lambda_{uv}$ and
$\Lambda_{ir}$  dependence of  the  integral (\ref{eq35}).   The sum  is
equivalent to the finite part  ( in the limit $\varepsilon $$\rightarrow
$0) of the dimensionally regularized integral.

Through  the  linearity  (\ref{eq13}-\ref{eq14}),  this  method  can  be
generalized to two  (or more) loop calculations. If  $q_\mu$ and $l_\mu$
are the  two loop virtual momenta,  then we can  rewrite the integration
variables      $\d^n     l=     l^{2(1+\varepsilon/2)}\d      l^2     \d
\Omega_{4+\varepsilon}/2$      and     $\d^n      q=      \d   
  q_{_\parallel}
q_{_\perp}^{2(1+\varepsilon/2)}\d    q_{_\perp}
    \d   \Omega_{3+\varepsilon}$.
$q_{_\parallel}$ is the component of $q_\mu$ parallel to $l_{\mu}$, 
$q_{_\perp}$ is
the total  length of  the components of  $q_\mu$ orthogonal  to $l_\mu$.
One then  defines two six  dimensional vectors $l_{\mu}$  and $q_{\mu}$,
analogously   to    (\ref{eq21}),   as   functions    of   five   angles
$\theta,\phi,\delta,\phi^\prime,\delta^\prime$,  two auxiliary variables
$x$ and $y$ (needed for the integrations $\d \Omega_{4+\varepsilon}$ and
$\d \Omega_{3+\varepsilon}$) and  the complex variables $l^2,\,
 q_{_\parallel},
\,  q_{_\perp}$.  Applying the  trick (\ref{eq15})-(\ref{eq19})  one obtains
the function  $w[ l^2,\,  q_{_\parallel}, \, q_{_\perp}]$.  
 Needless to  say, this
contains  dilogarithms  in  addition  to  some  logarithms.   From  this
guidelines,  one  gets  the  analogue of  the  (\ref{eq34}), for two
loop calculations
\cite{caravaglios:2000}.
\subsection*{Some  Numerical  Examples }  In  order  to  make clear  the
practical purposes  of these methods,  we discuss here some  very simple
examples.   Suppose that  we want  to  integrate \inlinebig{\label{eq37}
\int  {\d^n}l\frac{1}{{{(l+p)}^2}+{m^2}}}  with \inlinemio{{m^2}=2}  and
\inlinemio{{p^2}=1}.    Our  numerical   approach  must   reproduce  the
analytical     result      \inlinebig{\label{eq38}2     \frac{     {{\pi
}^2}}{\varepsilon }{m^2}  + {{\pi }^2} {m^2}  \log [{m^2}]+...}  Suppose
that we want to extract the finite part\footnote{ It is more instructive
to   discuss  the   finite   part,  since   the   integral  contour   in
(\ref{eq34bis})  is  trivial,  and   extracting  the  singular  part  is
straightforward. } \inlinemio{ {{\pi }^2} {m^2} \log [{m^2}]}.  We apply
the second method B described  above.  We use the five dimensional array
(\ref{eq21})  for $l_\mu$.  Then  the  integrand is  a  function of  $x,
t^2,\theta  ,\phi,\delta$.  The  finite  part  of  the  (\ref{eq37})  is
obtained computing the two integrals in $I_0$ (eq.(\ref{eq34})). For the
first  one  we  choose the  contour  of  integration  as in  figure 1:  the
numerical values of $\Lambda_{uv}$  and $\Lambda_{ir}$ have to be chosen
in  such  a  way  that  the  contour A+B+C  contains  all  the  physical
singularities  in  the  complex  $t^2$  plane.  Since  the  integral  is
infrared  convergent we  can take  the limit  $\Lambda_{ir}\rightarrow 0
$. Instead,  in the  ultraviolet region,  we can cut  the integral  at $
\Lambda_{uv}^2=9$.   We  also   must  set   $x=1$.  Then   the  integral
(\ref{eq35})  becomes \inlinebig{\label{eq39} P_1=\int_{0}^{{9}}  \d t^2
\Mvariable{\d \Omega_4  } \frac{1}{2} \frac{t^2}{{t^2}+2  t \cos [\theta
]+3}= 51.929.   } One can recognize  in the integral  above the ordinary
four dimensional integration with a cut-off $ \Lambda_{uv}^2$.  Then the
integration in  \inlinemio{{{\Mvariable{\d t }}^2}} follows  the path B
\inlinebig{\label{eq40}P_2=    \frac{1}{2   \pi    i}\oint_B    \d   t^2
\Mvariable{\d  \Omega_4 } \log  [-t^2] \frac{1}{2}  \frac{t^2}{{t^2}+2 t
\cos [\theta  ]+3}= -43.1815.   } This completes  the first  integral in
$I_0$.   The  second  one  in  eq.(\ref{eq34}) contains  a  non  trivial
integration  in the  variable $x$,  which  must be  performed using  the
prescription (\ref{eq26}) .   The contour of integration in  $\d t^2$ is
rather simple: there is no logarithm,  and no cut in the real axis; then
we do  not need to follow the  path A. There is  no infrared singularity
and also the  path C vanishes.  The only  non trivial contribution comes
from the path  B, a circle of radius $  \Lambda_{uv}^2$.  This gives \be
\label{eq41}  P_3= \frac{1}{\Mvariable{i  2  \pi }}\int  _{0}^{1}\oint_B
\frac{1}{{t^2}+2   x   t  \cos   [\theta   ]+3}  {t^2}   {{\Mvariable{\d
t}}^2}{{\bigg(\frac{{x^3}}{    1-{x^2}}    \bigg)}_+}\Mvariable{\d    x}
{{\Mvariable{\d \Omega }}_{4 }}= 4.935.  \ee Note in the denominator the
appearance of the variable $x$, which  comes out when we take the scalar
product $l\cdot p$ with $l_\mu $ from eq.(\ref{eq21}) and $p^\mu$ in the
direction  $\mu=1$.    The  sum   of  the  three   contributions  yields
\inlinebig{\label{eq42}
I_0\big[\frac{1}{{{(l+p)}^2}+{m^2}}\big]=P_1+P_2+P_3=       13.6825\simeq
{{\pi }^2} {m^2} \log [{m^2}],} in perfect agreement with the analytical
expression.   The method  can  also be  applied  for infrared  divergent
integrals.        The        integral        \inlinebig{\label{eq43}\int
{\d^n}l\frac{1}{l^6}\frac{1}{{{(l+p)}^2}+{m^2}}}      is     ultraviolet
convergent and infrared  divergent. Thus we choose $\Lambda_{uv}=\infty$
and  $\Lambda_{ir}=1/4$.    We  get  \inlinebig{\label{eq44}{P_1}  =\int
_{1/4}^{\infty  }  {t^2}{{\Mvariable{\d t}}^2}  \frac{1}{2}\Mvariable{\d
\Omega  }\frac{1}{t^2+2 t \cos[\theta]  +3} \frac{1}{t^6}  =10.843,} \ba
\label{eq45}P_2&=&  \frac{1}{2   \pi  i}\oint_C  \d   t^2  \Mvariable{\d
\Omega_4 }  \log [-t^2] \frac{1}{2 t^4} \frac{1}{{t^2}+2  t \cos [\theta
]+3}=  -12.1255 \nn  \\  \nn  & &  \\  \nn &  &  \\ \label{eq46}  P_3&=&
\frac{1}{\Mvariable{i 2 \pi  }}\int _{0}^{1}\oint_B \frac{1}{{t^2}+2 x t
\cos  [\theta  ]+3}  {t^2}  {{\Mvariable{\d  t}}^2}{{\bigg(\frac{{x^3}}{
1-{x^2}}  \bigg)}_+}\Mvariable{\d x}  {{\Mvariable{\d  \Omega }}_{4  }}=
-0.1825.  \\   \nn  \ea   The   sum  of   the   three  integrals   gives
\inlinebig{\label{eq47}I_0\big[\frac{1}{l^6}\frac{1}{{{(l+p)}^2}+{m^2}}\big]
={P_1}+{P_2}+{P_3}=-1.465,}   very  close  to   the  exact   result  \be
\label{eq48}       -\frac{1}{27}       {{\pi      }^2}       \Big(1+\log
\big[\frac{81}{4}\big]\Big).  \ee

\section*{Conclusions}  In   the  dimensional  regularization   of  loop
integrals, usually one defines the  integral as a function of the number
of dimensions in a region of $n$ where the integral is convergent.  Then
one  makes  an analytic  continuation  to  obtain  a definition  of  the
integral also in  regions of $n$ where the  integral is divergent.  This
procedure is  clear and unambiguous, however  it can only  be applied in
pure analytic (and symbolic)  calculations.  It cannot be converted into
an  obvious numerical  procedure.  One  is obliged  to  perform symbolic
calculations, which  sometimes becomes lengthy  and/or unaffordable.  In
this paper we have shown how  to set up a different approach, where each
coefficient  the  Laurent  expansion   in  powers  of  $\varepsilon$  in
eq.  (\ref{eq1}) can  be written  in  terms of  concrete and  convergent
integrals (well) defined in a five dimensional space (see eq.(\ref{eq8})
and  eqs.(\ref{eq33}),(\ref{eq34bis})  and  (\ref{eq34})  .  Instead  of
abstract $n$ dimensional vectors we  have to deal with ``concrete'' five
dimensional vectors, and we have  to perform integrations over a compact
space.  This formulation has the advantage to allow us the evaluation of
all integrals  through pure numerical methods.  Clearly  further work is
needed, to  generalize the above result  to more loops and  to prove the
efficiency of this technique in realistic physical problems.  However we
believe  that the  simplicity of  the numerical  approach makes  it very
promising.

\section*{Acknowledgements}  I would  like to  thank P.  Nason  for very
helpful discussions,  and Prof. R.  Ferrari and the  Theoretical Physics
Department   of    Milano   University   for    its   kind   hospitality.


\begin{thebibliography}{99}      

\bibitem{'tHooft:1972fi}
G.~'t Hooft and M.~Veltman,
Nucl.\ Phys.\  {\bf B44} (1972) 189.
\bibitem{Mangano:1991by}   M.~L.~Mangano   and
S.~J.~Parke, 
Phys.\ Rept.\
{\bf    200},    301     (1991).    




%
%

\bibitem{Mangano:1988xk}                                           
  M.L. Mangano, S.J. Parke and Z. Xu, Proceedings of Les Rencontres de
Physique de la Vallee d'Aoste, (La Thuile, Italy, 1987),
ed. M. Greco (Edition Frontieres), p.513;\\                                    
F.A. Berends and W.T. Giele, Nucl. Phys. {\bf B294}, 700 (1987); \\
M.~Mangano, S.~Parke and Z.~Xu,
Nucl.\ Phys.\  {\bf B298}, 653 (1988).
\bibitem{Mangano:1988kp}
M.~Mangano and S.~J.~Parke,
Nucl.\ Phys.\  {\bf B299}, 673 (1988).

%
\bibitem{Mangano:1988kk}
M.~Mangano,
Nucl.\ Phys.\  {\bf B309}, 461 (1988).


%
%
%
\bibitem{Parke:1986gb}
S.~J.~Parke and T.~R.~Taylor,
Phys.\ Rev.\ Lett.\  {\bf 56}, 2459 (1986).

\bibitem{DeCausmaecker:1982bg}
P.~De Causmaecker, R.~Gastmans, W.~Troost and T.~T.~Wu,
Nucl.\ Phys.\  {\bf B206}, 53 (1982).
%
\bibitem{Kleiss:1985yh}
R.~Kleiss and W.~J.~Stirling,
Nucl.\ Phys.\  {\bf B262}, 235 (1985).
\bibitem{Xu:1987xb}
Z.~Xu, D.~Zhang and L.~Chang,
Nucl.\ Phys.\  {\bf B291}, 392 (1987).



%
%


\bibitem{Berends:1988me}
F.~A.~Berends and W.~T.~Giele,
Nucl.\ Phys.\  {\bf B306}, 759 (1988);
F.~A.~Berends and W.~T.~Giele,
Nucl.\ Phys.\  {\bf B313} (1989) 595.


%


\bibitem{Caravaglios:1995cd}
F.~Caravaglios and M.~Moretti,
Phys.\ Lett.\  {\bf B358}, 332 (1995).



%
%

\bibitem{Kunszt:1988it}
Z.~Kunszt and W.~J.~Stirling,
Phys.\ Rev.\  {\bf D37}, 2439 (1988).


\bibitem{Maxwell:1989wx}
C.~J.~Maxwell,
Nucl.\ Phys.\  {\bf B316}, 321 (1989).

\bibitem{Mangano:1989rp}
M.~Mangano and S.~Parke,
Phys.\ Rev.\  {\bf D39}, 758 (1989).


%
%


\bibitem{Berends:1990hf}
F.~A.~Berends, W.~T.~Giele and H.~Kuijf,
Nucl.\ Phys.\  {\bf B333}, 120 (1990).




\bibitem{Kleiss:1989ne}
R.~Kleiss and H.~Kuijf,
Nucl.\ Phys.\  {\bf B312}, 616 (1989).




%
%
%
%

\bibitem{Passarino:1979jh}
G.~Passarino and M.~Veltman,
Nucl.\ Phys.\  {\bf B160}, 151 (1979).


\bibitem{Bern:1991cu}
Z.~Bern and D.~A.~Kosower,
Phys.\ Rev.\ Lett.\  {\bf 66}, 1669 (1991).
Z.~Bern and D.~A.~Kosower,
Nucl.\ Phys.\  {\bf B362}, 389 (1991).
Z.~Bern and D.~A.~Kosower,
Nucl.\ Phys.\  {\bf B379}, 451 (1992).
Z.~Bern and D.~C.~Dunbar,
Nucl.\ Phys.\  {\bf B379}, 562 (1992).

\bibitem{Bern:1994qk}
Z.~Bern, G.~Chalmers, L.~Dixon and D.~A.~Kosower,
Phys.\ Rev.\ Lett.\  {\bf 72}, 2134 (1994).


\bibitem{Bern:1996db}
Z.~Bern and A.~G.~Morgan,
Nucl.\ Phys.\  {\bf B467}, 479 (1996).

\bibitem{Bern:1996je}
Z.~Bern, L.~Dixon and D.~A.~Kosower,
Ann.\ Rev.\ Nucl.\ Part.\ Sci.\  {\bf 46}, 109 (1996).
%
%
%
\bibitem{davydychev:1999}
A.I. Davydychev, V.A. Smirnov  Nucl. Phys. {\bf B554}(1999)391.
\bibitem{fleischer:1999}
J. Fleischer, M.Yu. Kalmykov,  Phys. Lett. {\bf B470}(1999)168.

\bibitem{davydychev:1997}
A.I.~Davydychev, Acta Phys. Pol. {\bf 28}(1997)841 and references therein.
\bibitem{davydychev:1996}
A.I.~Davydychev, J.B.~Tausk Nucl. Phys. {\bf B465}(1996)507.
\bibitem{brucher:1997}
L.~Br\"ucher, J.~Franzkowski, A.~Fink and D.~Kreimer, Acta Phys. Pol.
 {\bf 28}(1997)835. 

%
\bibitem{Draggiotis:1998gr}
P.~Draggiotis, R.~H.~Kleiss and C.~G.~Papadopoulos,
Phys.\ Lett.\  {\bf B439}, 157 (1998).
\bibitem{caravaglios:1999}
F.~Caravaglios, M.L.~Mangano, M.~Moretti and R.~Pittau,
 Nucl. Phys. {\bf B539}, 215 (1999).

    
\bibitem{Caravaglios:1997nq}
F.~Caravaglios and M.~Moretti,
Z.\ Phys.\  {\bf C74}, 291 (1997).
\bibitem{Moretti:1997nv}
M.~Moretti,
Nucl.\ Phys.\  {\bf B484}, 3 (1997).
\bibitem{Montagna:1998dc}
G.~Montagna, M.~Moretti, O.~Nicrosini and F.~Piccinini,
Eur.\ Phys.\ J.\  {\bf C2}, 483 (1998).


\bibitem{Soper:1999xk}
D.~E.~Soper,
hep-ph/9910292.

\bibitem{caravaglios:2000}
F.~Caravaglios, work in progress.

\end{thebibliography}
\end{document}